 \documentclass[aps,pre,preprint,superscriptaddress,showpacs]{revtex4}%

\usepackage{amsmath}
\usepackage{amssymb}
\usepackage[dvips]{graphicx}
\usepackage{colordvi}
\usepackage{latexsym}
\usepackage{amsfonts}
\usepackage{times}

\newcommand{\mk}[1]{[[#1]]}

\newcommand{\be}{\begin{equation}}
\newcommand{\ee}{\end{equation}}
\newcommand{\bea}{\begin{eqnarray}}
\newcommand{\eea}{\end{eqnarray}}
\newcommand{\beas}[1]{\begin{subequations}\label{#1}\bea}
\newcommand{\eeas}{\eea\end{subequations}}
\newcommand{\width}{12cm}
\newcommand{\IM}[1]{{\bf I}^{(#1)}}
\newcommand{\RE}[1]{{\bf R}^{(#1)}}
\newcommand{\MAT}[3]{\left(\begin{array}{ccc} #1 \\ #2 \\ #3
\end{array}\right )}

\newcommand{\MR}{\Re({\bf M})}
\newcommand{\MI}{\Im({\bf M})}
\newcommand{\MRPRIME}{\Re({\bf M}')}
\newcommand{\MIPRIME}{\Im({\bf M}')}

\renewcommand{\Re}{ {{\rm Re}}}
\renewcommand{\Im}{ {{\rm Im}}}

\begin{document}

\title{From hyperbolic regularization to exact  hydrodynamics\\
  {for linearized Grad's equations}}

\author{Matteo Colangeli\email{matteo.colangeli@mat.ethz.ch}}
\affiliation{ETH Z\"urich, Department of Materials, Polymer Physics,
CH-8093 Z\"urich, Switzerland}

\author{Iliya V. Karlin}
\affiliation{ETH Z\"urich, Aerothermochemistry and Combustion
Systems Lab, CH-8092 Z\"urich, Switzerland}

\author{Martin Kr\"oger\homepage{www.complexfluids.ethz.ch}}
\affiliation{ETH Z\"urich, Department of Materials, Polymer Physics, and Materials Research Center,
CH-8093 Z\"urich, Switzerland}



\date{\today}

\begin{abstract}
Inspired by a recent hyperbolic regularization of Burnett's
hydrodynamic equations [A. Bobylev, J. Stat. Phys. {\bf 124}, 371
(2006)], we introduce a method to derive hyperbolic equations of
linear hydrodynamics to any desired accuracy in Knudsen number.
The approach is based on a dynamic invariance principle which
derives exact constitutive relations for the stress tensor and
heat flux, and a transformation which renders the exact equations
of hydrodynamics hyperbolic and stable. The method is described in
detail for a simple kinetic model - a thirteen moment Grad system.
\end{abstract}

\maketitle

\section{Introduction}

{Derivation of hydrodynamics from a microscopic description is the
classical problem of physical kinetics. The Chapman-Enskog method
\cite{chapman} derives the solution from the Boltzmann equation in a
form of a series in powers of Knudsen number $\varepsilon$, where
$\varepsilon$ is a ratio between the mean free path of a particle
and the scale of variations of hydrodynamic fields. The
Chapman-Enskog solution leads to a formal expansion of stress tensor
and of heat flux vector in balance equations for density, momentum,
and energy. Retaining the first order term ($\varepsilon$) in the
latter expansions, we come to the Navier--Stokes equations, while
next-order corrections are known as the Burnett \cite{burnett}
($\varepsilon^2 $) and the super-Burnett ($\epsilon^3 $) corrections
\cite{chapman}. It has long been conjectured that the inclusion of
higher-order terms in the constitutive relations for the stress and
heat flux may improve the predictive capabilities of hydrodynamics
formulations in the continuum­-transition regime where Navier-Stokes
equations fail.

However, as it was first demonstrated by Bobylev for Maxwell's
molecules \cite{bobylev1}, even in the simplest case
(one-dimensional linear deviation from global equilibrium), the
Burnett and the super-Burnett hydrodynamics violate the basic
physics behind the Boltzmann equation. Namely, sufficiently short
acoustic waves are increasing with time instead of decaying.
Bobylev's instability has been also studied by Uribe \emph{et al.}
\cite{uvgc} for hard sphere molecules.
 This instability contradicts the
$H$-theorem, since all near-equilibrium perturbations must decay.
This creates difficulties for an extension of hydrodynamics, as
derived from a microscopic description, into a highly
non-equilibrium domain where the Navier-Stokes approximation is
inapplicable. For example, higher-order systems of hydrodynamic
equations afforded a better description in certain situations such
as shock structures on coarse grids, but are prone to small
wavelength instabilities when grids are refined. Successes and
drawbacks of the Burnett computations and a family of extended
Burnett equations aimed at achieving entropy-consistent behavior
of the equations have been recently summarized in
\cite{Balakrishnan}.}

The failure of the CE expansion does not lie in the method itself,
but in its truncation to lower order levels. {This question was
studied in some detail for a class of
simple kinetic models - Grad's moment systems \cite{Grad} - in
Refs. \cite{Ka1996,Ka1997a,Ka1997b,Ka2000,Ka2002,Ka2005}. In these
works, the Chapman-Enskog expansion was summed up exactly which
revealed stability of exact hydrodynamics in contrast to its
finite-order approximations. Alternative ways of approximating the
Chapman-Enskog solution have been also suggested.

Very recently, Bobylev suggested a different viewpoint on the
problem of Burnett's hydrodynamics \cite{Bo2006}. Namely,
violation of hyperbolicity can be seen as a source of instability.
We remind that Boltzmann's and Grad's equations are hyperbolic and
stable due to corresponding $H$-theorems. However, the Burnett
hydrodynamics is not hyperbolic which leads to no $H$-theorem.
Bobylev \cite{Bo2006} suggested to stipulate hyperbolization of
Burnett's equations which can also be considered as a change of
variables. In this way hyperbolically regularized Burnett's
equations admit the $H$-theorem (in the linear case, at least) and
stability is restored.

The aim of this paper is to study the issue of hyperbolicity of
higher-order hydrodynamics in the case where the Chapman-Enskog
solution can be found exactly. As a starting point, we consider
Grad's moment system, linearized at the equilibrium, and assuming
unidirectional flow conditions (the 1D13M system, according to
\cite{Ka2005}). While simple enough, this model is nontrivial for
three reasons:
\begin{itemize} \item Application of the Chapman-Enskog method
leads to a rather involved nonlinear recurrent relations for the
coefficients of the expansion; \item The Burnett approximation
derived from Grad's moment system is identical to the one derived
from the Boltzmann equation for Maxwell's molecules and thus
violates hyperbolicity and exhibits Bobylev's instability
\cite{bobylev1}; \item Even though the exact hydrodynamics can be
derived following the lines of Refs.
\cite{Ka1996,Ka1997a,Ka1997b,Ka2000,Ka2002,Ka2005}, and is stable,
the question remains whether or not this exact hydrodynamics is
manifestly hyperbolic.
\end{itemize}}

The paper is organized as follows: In Sec.~\ref{sec:1}, we derive
exact hydrodynamics from the linearized 1D13M Grad's system.
Derivation closely follows \cite{Ka2002}, and is based on
application of a dynamic invariance principle which is equivalent
to exact summation of the Chapman-Enskog expansion. A critical
value of the Knudsen number is found beyond which a closed system
of equations for the locally conserved fields ceases to exist. In
Sec.~\ref{sec:2} we find a class of transformations through which
exact equations of hydrodynamics can be put in a hyperbolic form,
thereby answering in affirmative the above question. We also
analyze how such transformations affect the dissipative nature of
the equations. In Sec.~\ref{sec:3}, we analyze and compare with
conventional and earlier  approximate solutions provided by (i)
the Newton iteration method (Appendix~\ref{newton}) and (ii)
Bobylev's hyperbolic regularization of Burnett's equations
(Appendix~\ref{app:boby}) which turns out to be a special case of
a more general results presented here. Finally, conclusions are
provided in Sec.~\ref{sec:conclusions}.

\section{Hydrodynamics from the linearized Grad system}
\label{sec:1}

\subsection{Chapman-Enskog method and Bobylev's instability of Burnett's hydrodynamics}

Point of departure is the linearized Grad's thirteen-moment system
in one spatial variable $x$:
\begin{eqnarray}
  \partial_{ t} \rho  & = & -\partial_{x}u, \label{1} \nonumber \\
  \partial_{t}  u     & = & -\partial_{x}\rho-\partial_{x}T-\partial_{x}\sigma, \nonumber \\
  \partial_{t}  T     & = & -\frac{2}{3}\partial_{x}u-\frac{2}{3}\partial_{x}q,\label{1b} \\
  \partial_{t} \sigma & = & -\frac{4}{3}\partial_{x}u-\frac{8}{15}\partial_{x}q-
   {\frac{1}{\epsilon}}\sigma, \nonumber \\
  \partial_{t}  q     & = & -\frac{5}{2}\partial_{x}T-\partial_{x}\sigma-\frac{2}{3 {\epsilon}}q.
  \nonumber
\end{eqnarray}
Here $\rho(x,t)$, $u(x,t)$ and $T(x,t)$ are the reduced deviations
of density, average velocity and temperature from their
equilibrium values, respectively, and $\sigma(x,t)$ and $q(x,t)$
are reduced $xx$-component of the nonequilibrium stress tensor and
heat flux, respectively.  {Moreover, $\varepsilon>0$ has a meaning
of the Knudsen number. The latter is given by the ratio between
the mean free path $\lambda$ and a characteristic dimension of the
system $L$ and is the smallness parameter in the Chapman-Enskog
method \cite{chapman}. The magnitude of the Knudsen number
determines the appropriate gas dynamic regime \cite{mkmh}. In a
sequel, we use rescaled variables $t'=\varepsilon t$ and
$x'=\varepsilon x$ and omit prime to simplify notation.}

The system (\ref{1b}) provides the time evolution equations for a
set of hydrodynamic (locally conserved) fields $[\rho,u,T]$
coupled to the nonhydrodynamic fields $\sigma$ and $q$. The goal
is to reduce the number of equations in (\ref{1b}) and to arrive
at a closed system of three equations for the hydrodynamic fields
only. Thanks to linearity of the system (\ref{1b}) it proves
convenient to turn into the reciprocal space, and seek for
solutions of the form $\zeta=\zeta_{k}\exp(\omega t+ikx)$, where
$\zeta$ is a generic function $\rho, u, T, \sigma, q$, and where
$k$ is a real valued wavenumber.

Application of the Chapman-Enskog (CE) method to the reduction of
the system (\ref{1b}) results in the following series expansion of
the nonhydrodynamic variables into the powers of $k$:
\begin{equation}
\sigma_{k}=\sum_{n=0}^{\infty}\sigma_{k}^{(n)},\
q_{k}=\sum_{n=0}^{\infty}q_{k}^{(n)},
\end{equation}
where the coefficients $\sigma_{k}^{(n)}$ and $q_{k}^{(n)}$ are of
order $\sigma_{k}^{(n)}\sim k^{n+1}$, $q_{k}^{(n)}\sim k^{n+1}$,
and are obtained from a recurrence procedure:
\begin{eqnarray}
\sigma_{k}^{(n)}&=&-\left\{\sum_{m=0}^{n-1}\partial_{t}^{(m)}\sigma_{k}^{(n-1-m)}+\frac{8}{15}ikq_{k}^{(n-1)}\right\}, \nonumber \\
q_{k}^{(n)}&=&-\left\{\sum_{m=0}^{n-1}\partial_{t}^{(m)}q_{k}^{(n-1-m)}+ik\sigma_{k}^{(n-1)}\right\},
\label{eq6}
\end{eqnarray}
and where the CE operators $\partial_{t}^{(m)}$ act on the
hydrodynamic fields as follows:
\begin{eqnarray}
\partial_{t}^{(m)}\rho_{k} &=&
\left\{
\begin{array}{ll} -iku_{k} & , m=0
\\ 0 &,  m\geq 1
\end{array}
\right.,
\nonumber \\
\partial_{t}^{(m)}u_{k} &=&
\left\{
\begin{array}{ll}
 -ik(\rho_{k}+T_{k}) & , m=0 \\
 -ik\sigma_{k}^{(m-1)} & , m\geq1
\end{array}
\right.,
\nonumber \\
\partial_{t}^{(m)}T_{k} &=&
\left\{
\begin{array}{ll}
-\frac{2}{3}iku_{k} & , m=0 \\
-\frac{2}{3}ikq_{k}^{(m-1)} & , m\geq1
\end{array}
\right.. \label{CEoperators}
\end{eqnarray}
It can be proven that functions $\sigma_{k}$ and $q_{k}$ have the
following structure, for all $n=0,1,..$,
\begin{eqnarray}
\sigma_{k}^{(2n)}&=&a_{n}(-k^2)^niku_{k,}, \nonumber \\
\sigma_{k}^{(2n+1)}&=&b_{n}(-k^2)^{n+1}\rho_{k}+c_{n}(-k^2)^{n+1}T_{k}, \nonumber \\
q_{k}^{(2n)}&=&x_{n}(-k^2)^{n}ik\rho_{k}+y_{n}(-k^2)^{n}ikT_{k}, \nonumber \\
q_{k}^{(2n+1)}&=&z_{n}(-k^2)^{n+1}u_{k}, \label{Structure1D13M}
\end{eqnarray}
where $a_n,\dots,z_n$ are numerical coefficients to be determined.
Note the altering structure of expansion coefficients of odd and
even orders. Substituting (\ref{Structure1D13M}) into (\ref{eq6})
and (\ref{CEoperators}), the CE method casts into a recurrence
equations in terms of the coefficients $a_n,\dots,z_n$:

\begin{eqnarray}
a_{n+1}&=& b_{n}+\frac{2}{3}c_{n}+\frac{2}{3}\sum_{m=1}^{n}c_{n-m}z_{m-1}+\sum_{m=0}^{n}a_{n-m}a_{m}-\frac{8}{15}z_{n},\nonumber\\
b_{n+1}&=&
a_{n+1}+\sum_{m=0}^{n}a_{n-m}b_{m}+\frac{2}{3}\sum_{m=0}^{n}c_{n-m}x_{m}-\frac{8}{15}x_{n+1},\nonumber\\
c_{n+1}&=&
a_{n+1}+\sum_{m=0}^{n}a_{n-m}c_{m}+\frac{2}{3}\sum_{m=0}^{n}c_{n-m}y_{m}-\frac{8}{15}y_{n+1},\nonumber\\
x_{n+1}&=& z_{n}+\sum_{m=1}^{n}z_{n-m}b_{m-1}+\frac{2}{3}\sum_{m=0}^{n}y_{n-m}x_{m}-b_{n},\nonumber\\
y_{n+1}&=& z_{n}+\sum_{m=1}^{n}z_{n-m}c_{m-1}+\frac{2}{3}\sum_{m=0}^{n}y_{n-m}y_{m}-c_{n},\nonumber\\
z_{n+1}&=&
x_{n+1}+\frac{2}{3}y_{n+1}+\frac{2}{3}\sum_{m=0}^{n}y_{n-m}z_{m}+\sum_{m=0}^{n}z_{n-m}a_{m}-a_{n+1}.
\label{ReccurentSystem}
\end{eqnarray}
System (\ref{ReccurentSystem}) is solved recurrently subject to
the initial conditions,
 {\begin{equation} a_0=-\frac{4}{3},\ b_{0}=-\frac{4}{3},\
c_{0}=\frac{2}{3},\ x_{0}=0,\ y_{0}=-\frac{15}{4},\
z_{0}=-\frac{7}{4}.\label{InitialCondition}\end{equation}}
 {The initial conditions are obtained by evaluating the
functions $\sigma_k$ and $q_k$ up to the Burnett order (see Eq.
(\ref{burnett131}) below) and identifying the coefficients $a_0$,
$x_0$ and $y_0$ from the Navier-Stokes approximation and the
remaining coefficients $b_0$, $c_0$ and $z_0$ from the Burnett
correction.} Equation (\ref{ReccurentSystem}) defines six
functions,
\begin{equation}
A(k) = \sum_{n=0}^\infty a_n(-k^2)^n, \dots,
Z(k) = \sum_{n=0}^\infty z_n(-k^2)^n. \label{powser}
\end{equation}
Thus, the CE solution amounts to finding functions $A,\dots,Z$
(\ref{powser}). Note that by the nature of the CE recurrence
procedure, functions $A,\dots,Z$ (\ref{powser}) are real-valued
functions. Knowing $A,\dots,Z$ (\ref{powser}), we can express the
nonequilibrium stress tensor and heat flux as
\begin{eqnarray}
   \sigma_{k} & = & ikA(k)u_{k}-k^2B(k)\rho_{k}-k^2C(k)T_{k}, \label{sigma} \\
  \     q_{k} & = & ikX(k)\rho_{k}+ikY(k)T_{k}-k^2Z(k)u_{k}. \label{q}
\end{eqnarray}
Upon substituting these expressions into the Fourier-transformed
balance equations (\ref{1b}), we obtain the closed system of
hydrodynamic equations which is conveniently written in a vector
form,
\begin{equation}
\partial_t{\bf x}= {\bf M}{\bf x},\label{ClosedHydro}
\end{equation}
where ${\bf x}\equiv (\rho_k,u_k,T_k)$, and the matrix ${\bf M}$ has
the form
\begin{equation}
{\bf M}=\left(\begin{array}{ccc}
                        0 & -ik & 0 \\
                        -ik(1\!-\!k^2B) & k^2A & -ik(1\!-\!k^2C) \\
                        \frac{2}{3}k^2X & -\frac{2}{3}ik(1\!-\!k^2Z) & \frac{2}{3}k^2Y \\
                      \end{array}\right). \label{defM}
\end{equation}
With this, we find the dispersion relation for the hydrodynamic
modes $\omega(k)$ by solving the characteristic equation,

\begin{equation}
{\rm det}\left( {\bf M} - \omega{\bf I} \right)=0,
                           \label{disprel}
\end{equation}
with ${\bf I}$ the unit matrix.

\begin{figure}[t]
\includegraphics[width=\width]{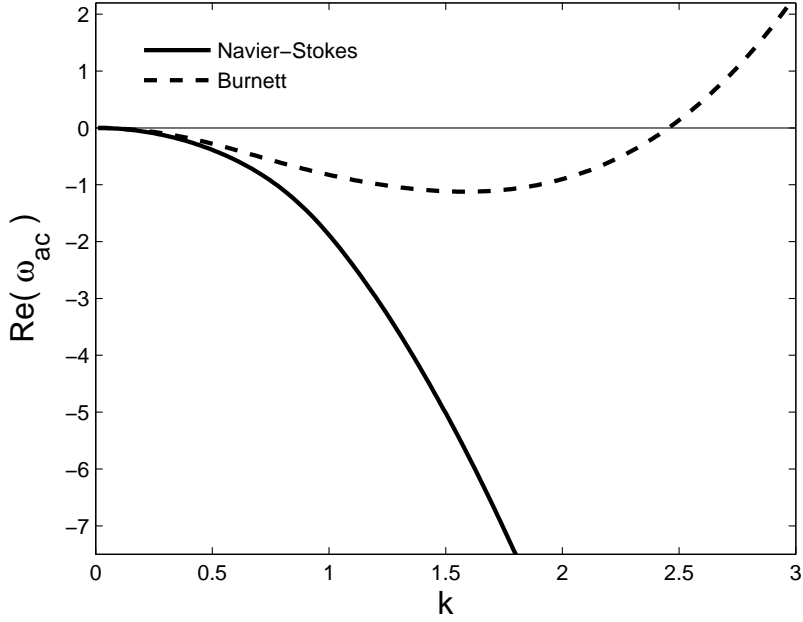}
\caption{Dispersion relation. Acoustic mode $\Re(\omega_{\rm ac})$
for Navier-Stokes and Burnett hydrodynamics.} \label{omegaplot}
\end{figure}

The standard application of the CE procedure is to approximate
functions $A,\dots,Z$ by polynomials with coefficients found from
the recurrence procedure (\ref{ReccurentSystem}). The first
non-vanishing contribution is the Newton-Fourier constitutive
relations,
\begin{equation}
\sigma_k^{(0)}=-\frac{4}{3}iku_k, q_k^{(0)}=-\frac{15}{4}ikT_k.
\end{equation}
which leads  to the Navier-Stokes-Fourier hydrodynamic equations.
Computing the coefficients $\sigma_k^{(1)}$ and $q_k^{(1)}$, we arrive at
the Burnett level:
\begin{eqnarray}
\label{burnett131} \sigma_{k} &=&-\frac{4}{3}iku_k +\frac{4}{3}k^2
\rho_k  -\frac{2}{3}k^2  T_k ,\\\nonumber
q_{k}&=&-\frac{15}{4}ikT_k +\frac{7}{4}k^2 u_k .
\end{eqnarray}
The Burnett approximation (\ref{burnett131}) coincides with that
obtained by Bobylev  \cite{bobylev1} from the Boltzmann equation for
Maxwell's molecules. Unlike the Navier-Stokes-Fourier approximation,
the Burnett constitutive relations (\ref{burnett131}) show
instability of the acoustic mode, see Fig.~\ref{omegaplot}.

Thus, the difficulty of the CE method consists in the way the
functions $A,\dots,Z$ are approximated, the standard polynomial
approximations lead to unstable hydrodynamic equations. We shall now
derive closed-form equations for these functions which amounts to
summing up the CE series exactly.

\subsection{Invariance Equations}

Summation of the CE series for the functions $A,\dots,Z$ can be
done directly from the recurrence relations
(\ref{ReccurentSystem}) following the lines of Ref.\
\cite{Ka1996}. Alternatively but completely equivalently, one can
make use of the  dynamic invariance principle (DIP) \cite{Ka2005}.
Here, the set of non-hydrodynamic moments $\{\sigma,q\}$ is still
thought in the form (\ref{sigma}) and (\ref{q}), but the method
makes no assumption about the power-series representation of the
functions $A,\dots,Z$. The time derivative of $\{\sigma,q\}$ can
be computed in two different ways. On the one hand, substituting
(\ref{sigma}) and (\ref{q}) into the moment system (\ref{1b}), we
find
\begin{eqnarray}
  \partial_{t} \sigma & = &
  -\frac{4}{3}iku_{k}-\frac{8}{15}ikq(X,Y,Z,k)-\sigma(A,B,C,k), \nonumber \\
  \partial_{t}  q     & = & -\frac{5}{2}ikT_{k}-ik
  \sigma(A,B,C,k)-\frac{2}{3}q(X,Y,Z,k).
  \nonumber
\end{eqnarray}
On the other hand, the time derivative of $\{\sigma,q\}$ can be
computed due to the closed hydrodynamic equations by chain rule:

\beas{chainrule}
  \partial_{t} \sigma & = & \frac{\partial\sigma}{\partial u_{k}}\,\partial_{t}u_{k}+
  \frac{\partial\sigma}{\partial \rho_{k}}\,\partial_{t}\rho_{k}+\frac{\partial\sigma}{\partial
  T_{k}}\,\partial_{t}T_{k},
 \\
  \partial_{t} q & = & \frac{\partial q}{\partial u_{k}}\,\partial_{t}u_{k}+
  \frac{\partial q}{\partial \rho_{k}}\,\partial_{t}\rho_{k}+\frac{\partial q}{\partial
  T_{k}}\,\partial_{t}T_{k}.
\eeas
Here, the derivatives $\partial_{t}u_{k}$ and
$\partial_{t}T_{k}$ are evaluated self-consistently using the
functions (\ref{sigma}) and (\ref{q}) in the right hand side of
(\ref{1b}). The DIP states that the two time derivatives coincide,
since the set $\{\sigma,q\}$ has to solve both the full Grad
system and the reduced system. This requirement implies a closed
set of equations, here referred as invariance equations (IE),
relating the six functions $A(k),..,Z(k)$:
\begin{eqnarray}
  -\frac{4}{3}-A-k^2(A^2\!+\!B\!-\frac{8Z}{15}\!+\!\frac{2C}{3})+\frac{2}{3}k^4CZ & = & 0, \nonumber \\
                                    \frac{8}{15}X+B-A+k^2AB+\frac{2}{3}k^2CX & = & 0, \nonumber \\
                                    \frac{8}{15}Y+C-A+k^2AC+\frac{2}{3}k^2CY & = & 0, \nonumber \\
                        A+\frac{2}{3}Z+k^2ZA-X-\frac{2}{3}Y+\frac{2}{3}k^2YZ & = & 0, \nonumber \\
                                  k^2B-\frac{2}{3}X-k^2Z+k^4ZB-\frac{2}{3}YX & = & 0, \nonumber \\
                 -\frac{5}{2}-\frac{2}{3}Y+k^2(C-Z)+k^4ZC-\frac{2}{3}k^2Y^2 & = & 0. \label{AZ}
\end{eqnarray}
The same equations can be derived upon summation of the CE
expansion. Equations (\ref{AZ}) are a convenient starting point
for evaluation of exact hydrodynamics.  {For $k=0$ one recovers
the initial conditions (\ref{InitialCondition}).}

\subsection{Exact hydrodynamic solutions}

\begin{figure}[t]
\includegraphics[width=\width]{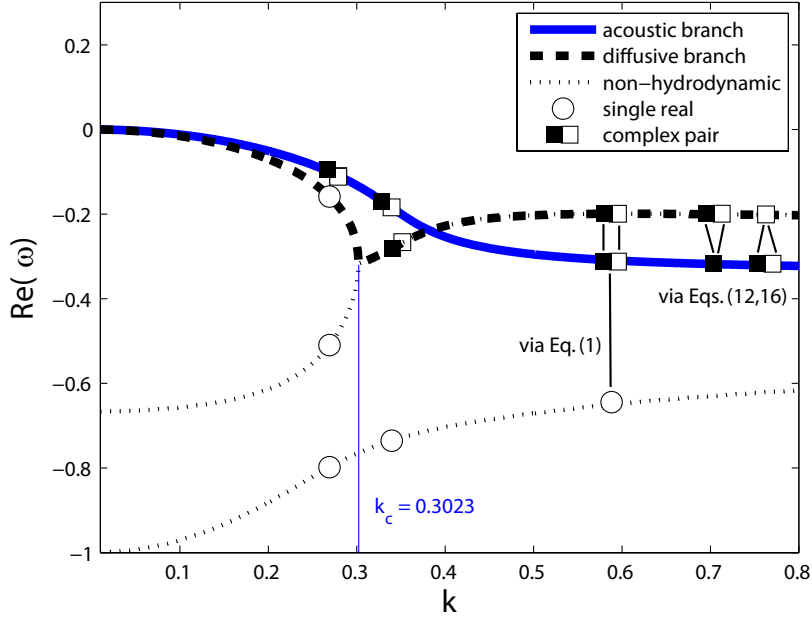}
\caption{(Color online) Dispersion relation for the linearized
1D13M Grad system (\ref{1}). The unique solution of hydrodynamical
modes obtained from (\ref{disprel}) with (\ref{AZ}) coincides with
the real parts of the modes of the original Grad system (the plot
also shows when pairs of conjugate complex roots appear), the
solution of the original system (\ref{1}) features five
$\omega$'s, while the exact solution of (\ref{disprel}) with
(\ref{AZ}) has three $\omega$'s for each $k$ and degenerated over
the hydrodynamic branches at $k\ge k_c$.} \label{mkfigover}
\end{figure}

The dispersion relation  $\omega(k)$ was found by simultaneously
solving numerically the invariance equations (\ref{AZ}) and the
characteristic equation (\ref{disprel}). The resulting
hydrodynamic spectrum consist of two modes, the acoustic mode,
$\omega_{\rm ac}(k)$, represented by two complex-conjugated roots
of (\ref{disprel}), and the real-valued diffusive heat mode,
$\omega_{\rm diff}(k)$. cf. Fig.~\ref{mkfigover}.

\begin{figure}[t]
\includegraphics[width=\width]{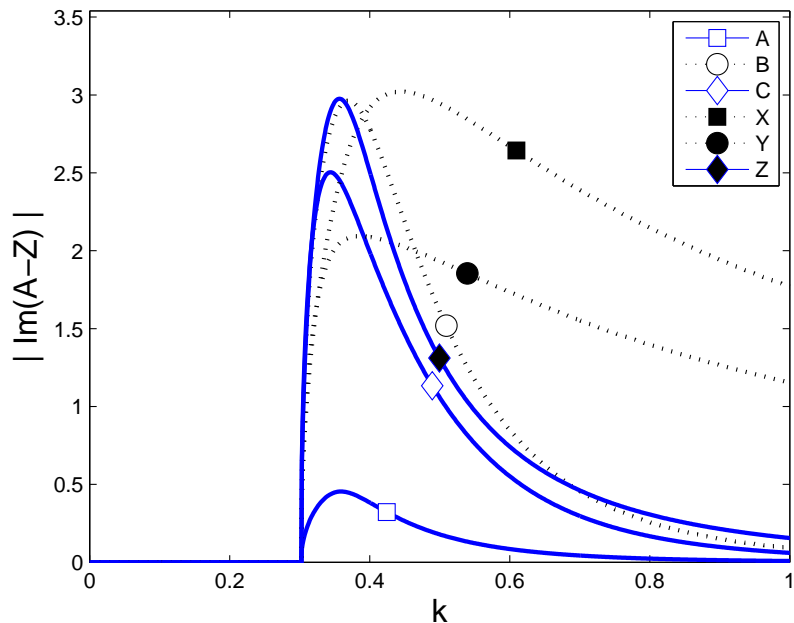}
\caption{(Color online) Imaginary parts of coefficients $A$ to $Z$
solving (\ref{AZ}). Shown is the unique solution leading to
hydrodynamic branches, cf. Fig.~\ref{mkfigover}, which is
symmetric about the real axis.} \label{mkfig4}
\end{figure}

\begin{figure}[t]
\includegraphics[width=\width]{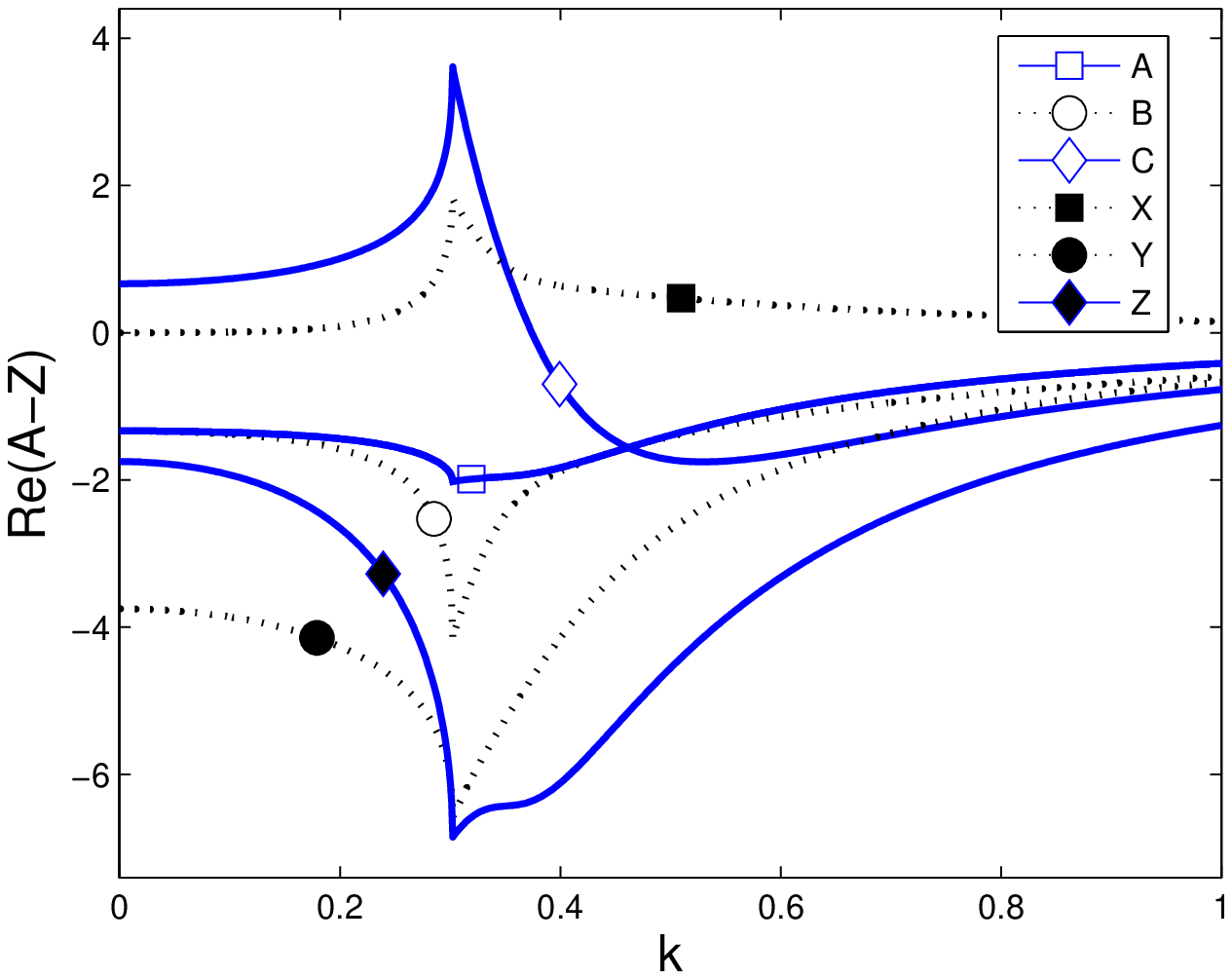}
\caption{(Color online) Real parts of complex-valued functions
$A,\dots,Z$ solving (\ref{AZ}).  {It is clearly visible that the
solution matches the initial condition
(\ref{InitialCondition}).}}\label{mkfig3}
\end{figure}

Among the many sets of solutions $\{A(k),...,Z(k)\}$ to the system
(\ref{AZ}), the relevant ones are continuous functions  with the
asymptotics: $\lim_{k\rightarrow0}\omega_{\rm hydr}=0$.
Remarkably, we find that the solution with this asymptotics is
unique, and represented by a pair of complex conjugated sets,
$[\mathcal{S},\mathcal{S}^{*}]$, shown in Figs.~\ref{mkfig4} and
\ref{mkfig3}. Note that a qualitative change of dynamics arises
when the diffusive mode couples with one of the two
non-hydrodynamical modes of Grad's system at a critical wave
number $k_{c}\approx0.3023$, which is the value where also the
Newton method diverges, cf. App.~(\ref{newton}). By the CE
perspective, the hydrodynamics of the diffusive mode stops at
$k_{c}$, since, after that point, it becomes a complex-valued
function coupled with the conjugated non-hydrodynamic mode, see
Fig.~\ref{mkfigover}. Essentially, for $k\ge k_{c}$, the CE method
does not recognize any longer the resulting diffusive branch as an
extension of a hydrodynamic branch. Also, the set of solutions
$[\mathcal{S},\mathcal{S}^{*}]$, real valued for $k\le k_{c}$,
continues upon a complex manifold, cf. Fig.~\ref{mkfig4}. We
notice that the occurrence of a pair of complex conjugated sets of
solution is very plausible due to symmetry reasons: inserting
$\mathcal{S}$ into the dispersion relation, we obtain a pair of
complex conjugated acoustic modes $[\omega_{\rm
ac}(\mathcal{S},k),\omega_{\rm ac}^{*}(\mathcal{S},k)]$ plus one
of the complex modes resulting from the extension of the diffusive
branch for $k\ge k_{c}$; whereas, through $\mathcal{S}^{*}$, we
obtain, symmetrically, the two latter conjugated modes, plus one
of the conjugated acoustic modes.

As a further evidence of this close coupling, we also notice the
occurrence of an intersection between the real parts of the
hydrodynamical modes $\Re(\omega_{\rm ac})$ and $\Re(\omega_{\rm
diff})$ after the critical point, at $k=k_{\rm coupl}\approx
0.383$. Therefore the message extracted  {from} the study of
Grad's system (\ref{1b}) is that the set of hydrodynamic equations
for $[\rho,u,T]$ provides, as expected, stable solutions, when
taking into account all the orders of CE expansion - which
corresponds to solving the system of invariance equations
(\ref{AZ}). And, there is no closed set of hydrodynamic equations
after $k_{c}$, even though the acoustic mode extends smoothly
beyond $k_{c}$, as is visible in Fig.~\ref{mkfigover}.

Thus, the exact hydrodynamics as derived by the summation of the CE
expansion (or, equivalently, from the invariance equations) extends
up to a finite critical value $k_{c}$. No stability violation
occurs, unlike in the finite-order truncations thereof. While we
have evaluated the functions $A,\dots,Z$ numerically, two questions
remained open:
\begin{itemize}
\item Is the (stable) exact hydrodynamics also hyperbolic?
\item
If so, how to retain hyperbolicity in the approximations?
\end{itemize}
In the next section we shall answer the first of these questions.

\section{Hyperbolic transformation for exact hydrodynamics}
\label{sec:2}

Equation (\ref{1b}) for the Fourier component vector ${\bf
x}\equiv (\rho_k,u_k,T_k)$ reads $\partial_t{\bf x} = {\bf M}{\bf
x}$ with ${\bf M}$ from (\ref{defM}). By explicitly re-introducing
the Knudsen number $\varepsilon$, i.e. by replacing $k$ by
$k\varepsilon$ in ${\bf M}$ and further distinguishing between the
real and imaginary matrix elements in $\bf M$, we can write
\bea
 \partial_{t} {\bf x} &=& [\MR-i\,\MI]{\bf x}, \label{shorta} \\
 \MR &=& \sum_{n=0}^\infty  (-1)^n \RE{n} \varepsilon^{2n+1}
 = \varepsilon \RE{0} - \varepsilon^3 \RE{1} + O(\varepsilon^5), \nonumber \\
 \MI &=& \sum_{n=0}^\infty  (-1)^n \IM{n} \varepsilon^{2n}
 = \IM{0} - \varepsilon^2\IM{1} + \varepsilon^4 \IM{2} - O(\varepsilon^6)  ,
\label{short} \eea rearranged such that the Knudsen number
expansion coefficients become visible. We find that the operators
$\MR$ (real part) and $\MI$ (imaginary part) involve the following
real-valued operators (for all $n\ge 0$ , i.e., with the
convention $a_{-1}=c_{-1}=z_{-1}\equiv 1$ and Kronecker symbol
$\delta$),
\begin{eqnarray}
 \IM{n} &=&  k^{2n+1} \MAT{0 & \delta_{n,0} & 0}{b_{n-1} & 0 & c_{n-1}}{ 0 & \frac{2}{3}z_{n-1} & 0}
 , \nonumber \\
 \RE{n} &=& k^{2n+2}\MAT{0 & 0 & 0}{0 & a_n & 0}{\frac{2}{3} x_n & 0 & \frac{2}{3} y_n}.
 \label{mats}
\end{eqnarray}


Equations of hydrodynamics (\ref{shorta}) are hyperbolic and
stable provided that we can find a transformation of hydrodynamic
fields such that
\begin{enumerate}
\item[(i)] $\MR$ and $\MI$ are both real and symmetric, and
\item[(ii)] $\MR$ has negative semidefinite eigenvalues.
\end{enumerate}
Therefore, we seek a transformation ${\bf z} = {\bf T}{\bf x}$
which produces a symmetric matrix ${\bf M}'={\bf T}{\bf M}{\bf
T}^{-1}$ and we wish to see if $\MRPRIME=\Re({\bf T}{\bf M}{\bf
T}^{-1})$ is negative semidefinite.
%
We consider the equations of exact hydrodynamics, i.e. equations
(\ref{shorta}) provided that functions $A,..,Z$ (\ref{mats}) are
solutions to the invariance equations (\ref{AZ}).
After a few algebra which we do not recapitulate here,
we obtain a particular transformation matrix ${\bf T}$
which solves the problem.
It is a member of a
whole class of effectively equivalent transformations, and can be
written as
\newcommand{\AB}[1]{[#1]}
\begin{eqnarray}
{\bf T} &=& \frac{1}{T_{uu}} \left(
\begin{array}{ccc}
 T_{\rho\rho} & 0 &
 T_{\rho T} \\
 0 & T_{uu} & 0 \\
 0 & 0 &
 T_{TT}
\end{array}\right),
\label{main}
\eea
with the nonvanishing components
\bea
 T_{\rho\rho} &=& \frac{T_{uu}^2}{\sqrt{3X+2 Y \mk{Z}}},
 \nonumber \\
 T_{uu} &=& \sqrt{X \mk{3 B- 2Z\mk{C}- 2C}+2 Y \mk{B}\mk{Z}},\nonumber \\
 T_{\rho T} &=& -\frac{3 \mk{C} X}{\sqrt{3X+2 Y \mk{Z}}}, \nonumber\\
 T_{TT} &=& \sqrt{3\mk{C} \left(Y\mk{B}-\mk{C} X\right)},
 \label{mainadd}
\eea
%
where we have introduced the following symbolic notation:
\begin{equation} \mk{\bullet} \equiv 1-(k\varepsilon)^2 \,\bullet\;.
\end{equation}
%
%
%
%
The transformation ${\bf T}$ (\ref{main}) symmetrizes $\bf M$ and
renders the system hyperbolic, as can be verified by straightforward
computation of  ${\bf M}'$ from (\ref{main}), (\ref{Mprimee}). We
further notice, that ${\bf T}$ contains only even powers of
$(k\varepsilon)$ because the same is true for the coefficients
$A$--$Z$.

Next we calculate the eigenvalues $\lambda_{1,2,3}$ of $\MRPRIME$
-- containing transport coefficients -- to obtain a remarkably
simple result:
\begin{equation}
\lambda_1=0, \;\;\;\lambda_2=k^2\varepsilon A, \;\;\;
\lambda_3=\frac{2}{3}k^2 \varepsilon Y. \label{18}
\end{equation}
From the analysis of the previous section, it follows that the
nontrivial eigenvalues (\ref{18}) are negative semidefinite for all
$k\varepsilon$ (see Fig.~\ref{mkfig3} which displays the exact
numerical solutions for $A$ and $Y$). Hence, the equation describing
hyperbolic hydrodynamics (also hyperbolic up to an arbitrarily
selected order $\varepsilon^{n}$, a feature to be used in the next
section) attains the form: \beas{final}
 \partial_{t} {\bf z} &=&\bf M'{\bf z}, \\
 {\bf M}' &=& {\bf T}\bf M{\bf
T}^{-1}\label{Mprimee} \eeas for the vector $ {\bf z}=\{
\tilde{\rho}_{k}, \tilde{u}_{k}, \tilde{T}_{k}\}={\bf T}{\bf x} $ of
transformed hydrodynamic variables, and where ${\bf M}'$ is
symmetric and has seminegative eigenvalues. To summarize, \beas{finalconditions}
  &&  \mbox{Hyperbolicity:}  \;\;\; ({\bf M}')^{T} = {\bf M}' , \\
  &&  \mbox{Dissipativity:}   \left\{ \begin{array}{lll}   {\rm Tr}[\MRPRIME]&\le &0 , \\
        {\rm det}[\MRPRIME]&\ge &0. \end{array}\right.
\eeas
%

Equation (\ref{final}) with (\ref{main}) and (\ref{defM})
{satisfying (\ref{finalconditions})} is the main result of this
paper. The occurrence of negative eigenvalues in the exact
solutions, together with the existence of a transformation ${\bf
T}$ which makes the equations hyperbolic, proves that exact
hydrodynamics (\ref{1b}), without approximations, is stable. In
the remainder of this paper we shall make use of the hyperbolicity
of exact hydrodynamics in order to establish approximate
hydrodynamic equations which retain this property.

\section{Lower order hyperbolic and stable hydrodynamics}
\label{sec:3}

\subsection{Approximations on the hyperbolic equations}
In applications, one is interested in using truncated hydrodynamic
equations by taking into account only lower orders of the Knudsen
number $\varepsilon$. In this case, the functions $A,...,Z$ are
replaced by their lower-order approximations, and they can be
generally written -- as shown already in Eq.~(\ref{powser}) -- as
polynomials truncated to an arbitrary order $n$. Their
coefficients are usually derived through the CE recurrence
equations, as outlined above. With the exact numerical solution at
hand, we can also find, at any given order of approximation, the
optimal interpolating functions $A,..,Z$ solving (\ref{AZ}), a
method we wish to recommend, and which has been worked out in
Tab.~\ref{tab1}. Exact hydrodynamics, as described by Grad's
system (\ref{1b}), terminates at $k_{c}$. In this regime we can
perform a Taylor expansion, up to any order $n$, upon the elements
of all the three matrices ${\bf T}$, ${\bf M}$, and ${\bf
T}^{-1}$. Thus, the approximations are done on the manifestly
hyperbolic equation (\ref{final}) in such a way as to retain
hyperbolicity and stability in each order of approximation. Is is
worthwhile noticing that the eigenvalues, upon approximating
Eq.~(\ref{final}) to a polynomial order $n$, transform in a
canonical manner:
\begin{equation}
  \lambda^{(n)}_1=0, \;\;\;\lambda^{(n)}_2=k^2\varepsilon
  [a_{0}+\sum_{m=1}^n a_m (k\varepsilon )^{m}],
\;\;\; \lambda^{(n)}_3=\frac{2}{3}k^2 \varepsilon
[y_{0}+\sum_{m=1}^{n}y_{m}(k\varepsilon )^{m}], \label{eigen}
\end{equation}
and, depending upon the polynomial coefficients, and in particular
depending on the sign of the highest order coefficients $a_n$,
$y_n$, the eigenvalues $\lambda_{2,3}$ diverge to $\pm\infty$ for
$k\varepsilon\rightarrow\infty$, but stay negative for $k\le k_c$,
if we use coefficients according to the method summarized in
Tab.~\ref{tab1}. We shall now consider a few examples of the
suggested procedure.

\subsection{Euler and Navier-Stokes equations}

For the zeroth-order term, $\MI=\IM{0}$ (Euler), the
transformation is, according to (\ref{mainadd}), given by a
diagonal matrix with entries $T_{\rho\rho}=T_{uu}=1$ and
$T_{TT}=\sqrt{3/2}$, all eigenvalues are identically zero. The first
order term, linear in the Knudsen number (Navier Stokes) is
obviously stable as well;
all eigenvalues are negative semidefinite
since $a_0=-4/3$ and $y_0=-15/4$ are both negative.

\subsection{Hyperbolic regularization for the Burnett level}

The Burnett equations are unstable without regularization. For
this level of description, second order in the Knudsen number
$\varepsilon$, with $\MI=\IM{0}-\varepsilon^2 \IM{1}$, upon
inserting the required exact solutions at vanishing wave number,
$a_0$, .., $z_0$ from (\ref{AZ}), cf.~Tab.~\ref{tab1}, into
(\ref{main}), (\ref{mainadd}), the transformation matrix achieving
a symmetric $\MIPRIME$ reads
\begin{equation}
 {\bf T} = \MAT{1+\frac{2}{3} (k\varepsilon)^2 & 0 & 0}
 {0 & 1  & 0}{0 & 0 &
 \sqrt{\frac{3}{2}} - \frac{29}{8\sqrt{6}}(k\varepsilon)^2 }. 
 \label{Torder2}
\end{equation}
%

This transformation coincides with the one derived by Bobylev's
hyperbolic regularization method \cite{Bo2006}, specified for the
present model (an alternate derivation which follows closely Ref.\
\cite{Bo2006} is given in Appendix~\ref{app:boby}).
Notice, that up to the Burnett
level
only the polynomial coefficients at vanishing wave number, listed in
the first row of Tab.~\ref{tab1}, enter
the transformation ${\bf T}$, which can be indirectly also inferred
from the eigenvalues, cf. Eq.~(\ref{eigen}).

\subsection{Beyond the Burnett level}

In Tab.~\ref{tab1}, we provide not only coefficients, but also
ranges of applicability for the given coefficients of $A-Z$ which
can be used if we extend the procedure to higher order. The optimal
coefficients
%
%
%
are provided by the least squares fit of the numerical data for
exact hydrodynamics, see Tab.~\ref{tab1}. Within the given ranges,
the eigenvalues of $\Re({\bf M}')$ are negative semidefinite, i.e.,
the spectrum of the acoustic mode $\omega_{\rm ac}(k)$ of the
corresponding hyperbolic hydrodynamic system is then stable for all
wavelengths.

\begin{table}
\begin{tabular}{cccrrrrrrr}
\hline\hline
 method & apply at  & $n$ & $a_n/n!$ & $b_n/n!$ & $c_n/n!$ & $x_n/n!$ & $y_n/n!$ & $z_n/n!$ \\ \hline
 0 & $k\varepsilon\le 0.03$ & 0 & -4/3 & -4/3 & 2/3 & 0 & -15/4 & -7/4 \\
 1 & $k\varepsilon\le 0.17$ &
  0 & -4/3 & -4/3 & 2/3 & 0 & -15/4 & -7/4 \\
 & & 1 & 1.132  &  2.536  &  -3.735  &  -0.716 &  5.873 &  9.953 \\
 2 & $k\varepsilon\le 0.25$ &
  0 & -4/3 & -4/3 & 2/3 & 0 & -15/4 & -7/4 \\
 & & 1 & 0.706  & 1.156  &  -2.500  & 0.309 &  4.652 &  7.053 \\
 & & 2 &-1.304  & -4.095   & 3.720  &  3.030 &  -3.741 &  -8.902 \\
 3 & $k\varepsilon\le 0.28$ &
  0 & -4/3 & -4/3 & 2/3 & 0 & -15/4 & -7/4 \\
 & & 1 & 1.104  &  3.042  & -4.055   & -1.123 &  5.903  & 9.718 \\
 & & 2 & 0.329  &  3.669  & -2.678  & -2.861 &   1.398  &  2.023 \\
 & & 3 &0.648  & 3.083  &  -2.540  &  -2.340 &  2.040  & 4.333 \\
\hline\hline
\end{tabular}
\caption{Polynomial coefficients introduced in (\ref{powser})
obtained from the exact numerical solution, cf. Fig.~\ref{mkfig3},
by requiring that deviations between exact and polynomial series
at given order of the method (first column) stay below 1\% (i.e.
would be invisible in the plot). We used the symmetrized functions
$A(k)+A(-k)$ over the whole real axes for $k$ when performing the
fits in order to enforce correct symmetry. This criterion
corresponds to a regularization procedure which produces stable
results up to the limit $(k\varepsilon)\le
(k\varepsilon)_c=0.3023$, as is easily verified, and leads to a
recommended range of (high precision) applicability of the method
(second column). For convenience we list faculty-rescaled series
coefficients. These coefficients are essentially the prefactors
for higher order correction terms in hydrodynamic equations and
can be used to study the intermediate Knudsen number regime $0\ll
k\varepsilon < (k\varepsilon)_c$. As described in the text part,
with a suitable transformation matrix ${\bf T}$ these choices lead
to very convenient hyperbolic differential equations for the
hydrodynamic fields ${\bf x}=(\rho,u,T)$. } \label{tab1}
\end{table}

\subsection{Application: Hyperbolic regularization for the super-Burnett level} \label{apphigher}

Finally, in order to present explicit illustration of the
approximation strategy presented in Sec.~\ref{sec:2}, we present
the equations on the next, Super-Burnett, level, which takes into
account terms up to the order $(k\varepsilon)^3$. The equations of
change for the transformed variables ${\bf z}$ read
\begin{eqnarray}
\partial_t {\bf z} &=& {\bf M}'{\bf z},
\label{sb1}
\end{eqnarray}
with a symmetric ${\bf M}'$,
\begin{eqnarray}
{\bf M}' &=& -ik\left\{ \MAT{0 & 1 & 0}{1 & 0 &
\sqrt{\frac{2}{3}}}{0 & \sqrt{\frac{2}{3}} & 0} +
(k\varepsilon)^2\MAT{0 & \frac{2(5+x_1)}{15} &
0}{\frac{2(5+x_1)}{15} & 0 & \frac{65-24x_1}{60\sqrt{6}} }{0 &
\frac{65-24x_1}{60\sqrt{6}} & 0} \right\} \nonumber \\ &&
-k^2\varepsilon\left\{
 \MAT{0 & 0 & 0}{0 & \frac{4}{3} & 0}{0 & 0 & \frac{5}{2}}
+(k\varepsilon)^2\MAT{0 & 0 & \sqrt{\frac{2}{3}} x_1}{0 & a_1 &
0}{\sqrt{\frac{2}{3}} x_1 & 0 & \frac{2}{3}y_1} \right\},
\label{oursuper}
\end{eqnarray}
and transformation
\begin{eqnarray}
 {\bf T} &=&
 \MAT{1 & 0 & 0}{0 & 1 & 0}{0 & 0 & \sqrt{\frac{3}{2}} }
 + (k\varepsilon)^2\MAT{\frac{2(5-x_1)}{15} & 0 &
 \frac{2x_1}{5}}{0 & 0 & 0}{0 & 0 & -\frac{145+24x_1}{40\sqrt{6}}},
 \label{Torder3}
\end{eqnarray}
where third order terms are not present because ${\bf T}$ is
symmetric in $k$. The Burnett level (\ref{Torder2}), where $x_1$
disappears, is immediately recovered from (\ref{Torder3}). The
hydrodynamic variables are restored using ${\bf x}={\bf
T}^{-1}{\bf z}$,
 with the inverse transformation (suitable at the super-Burnett level), which, due to
 our (arbitrarly) chosen normalization factor $T_{uu}$ in (\ref{main}) only
 slightly differs from
 ${\bf T}$:
\be
 {\bf T}^{-1} =
 \MAT{1 & 0 & 0}{0 & 1 & 0}{0 & 0 & \sqrt{\frac{2}{3}} }
 - (k\varepsilon)^2\MAT{\frac{2(5-x_1)}{15} & 0 &
 \frac{2x_1}{5}\sqrt{\frac{2}{3}}}{0 & 0 & 0}{0 & 0 & -\frac{145+24x_1}{60\sqrt{6}}}.
 \label{sb5}
\ee To complete the ``simulation algorithm'' using
(\ref{sb1})--(\ref{sb5}), we need numerical values for $x_1$ and
$y_1$, and an initial condition for ${\bf x}$, or ${\bf z}$. One
solves the hyperbolically stable system for ${\bf z}$, and finally
calculate ${\bf x}$ via ${\bf T}^{-1}$. Suitable values for the
coefficients are those given in Tab.~\ref{tab1} for method 1:
$y_1=5.873$ and $x_1=-0.716$, because higher order coefficients
such as $x_2$ do not enter. The equations of this section should
allow to study the regime $0\le k\varepsilon<0.17$ very
accurately. For the remaining regime, $0.17< k\varepsilon <
(k\varepsilon)_c$ the presented equations are also stable and
hyperbolic, but not as accurate. They are, by definition, more
accurate compared with the ones obtained using the recursion
method. The equations offered in this section serve as an example
on how to use our more general result, Eq.~(\ref{main}).

\section{Conclusions}
\label{sec:conclusions}

In this paper, we have considered derivation of hydrodynamics
for a simple model (\ref{1b}) for which -- as we have demonstrated -- all details can be
explicitly studied. The main finding is that the exact
hydrodynamic equations (summation of the Chapman-Enskog expansion
to all orders) are manifestly hyperbolic and stable. To the best of
our knowledge, this is the first complete answer of the kind. We
have also suggested a way of approximating the higher-order
hydrodynamic equations using accurate numerical solution of the
invariance equations and expansion of the transformation which
renders the system hyperbolic. The study supports the recent
suggestion of Bobylev on the hyperbolic regularization of
Burnett's approximation, and reduces to the latter in a special
case.

{We conclude this paper with a few comments on the possible
extensions of the present approach. (i) The technique of deriving
exact hydrodynamics and/or hyperbolic approximations thereof can
be readily applied to linearized Grad's systems with a larger
number of moments. In particular, we were able to extend the
present derivation to the three-dimensional thirteen moment
system, the results qualitatively agree with the one-dimensional
case considered above and will be reported separately. (ii) It is
also possible to apply the present techniques to derive exact
hydrodynamics from the dynamically corrected Grad's systems, first
introduced in \cite{Karlin98} and studied in some detail in
\cite{Struchtrup03}. The latter equations have arguably better
properties than the Grad's equations, especially in the moderate
Knudsen number regime where the linearized systems become relevant
to study of micro-flows. (iii) In this paper, we were addresing
the boundary conditions for neither the Grad's systems nor for the
higher-order hydrodynamic equations. As is well known, this
question remains essentially open for both. Therefore, a different
and intriguing field of applications of the present technique is
the recently established lattice Boltzmann hierarchy (LBH)
\cite{ShanHe98,K99,Ansumali02,Ansumali03,Ansumali05,Chikatamarla06,Chikatamarla06a,Ansumali07,Karlin07}.
Although the primitive variables in the LBH are populations of
carefully chosen discrete velocities, the LBH equations can be
cast into a  form of moment systems similar to Grad's equations.
The crucial advantage of the LBH above Grad's systems is that the
former is equipped with pertinent boundary conditions derived
directly from the Maxwell-Boltzmann theory \cite{Ansumali02}.
Recently, it has been demonstrated, both numerically and
analytically, that the LBH is capable of capturing such phenomena
as slip and nonlinear Knudsen layers \cite{Ansumali07,Karlin07}.
The present techniques can be applied for reducing higher-order
lattice Boltzmann models with advantage for the numerics. However,
this goes beyond the scope of this paper, interested reader is
directed to \cite{Chikatamarla06a,Ansumali07,Karlin07} for
details.}

\subsection*{Acknowledgment}

The authors thank Hans Christian \"Ottinger for very helpful
suggestions. M.K. acknowledges financial support through contracts
NMP3-CT-2005-016375 and FP6-2004-NMP-TI-4-033339 of the European
Community. I.V.K. gratefully acknowledges support by BFE Project
100862  and by CCEM-CH.

\begin{appendix}

\section{Newton iteration}
 \label{newton}

The analytical complexity of either the CE method or the
invariance equations is overwhelming when we regard systems other
than the linearized Grad system, such as the Boltzmann equation.
Approximate solutions are, then, the only feasible approach. In
this section we shall describe application of the Newton iteration
method to the invariance equations. We used Newton's method, cf.
Fig.~\ref{mkfignewton}, to solve iteratively the equations
(\ref{1b}), taking, as initial condition, the Euler approximation
(corresponding to a non-dissipative hydrodynamics:
$A_{0}=...=Z_{0}=0$), which leads, after the first iteration, to
the same result achievable, alternatively, through a technique of
partial summation \cite{Ka2005} of the CE expansion: essentially,
a sort of regularized Burnett approximation. It is seen in
Fig.~\ref{mkfignewton} that Newton iterations converge rapidly to
the exact hydrodynamics in the domain of its validity, $k\le k_c$.

\begin{figure}[t]
\includegraphics[width=\width]{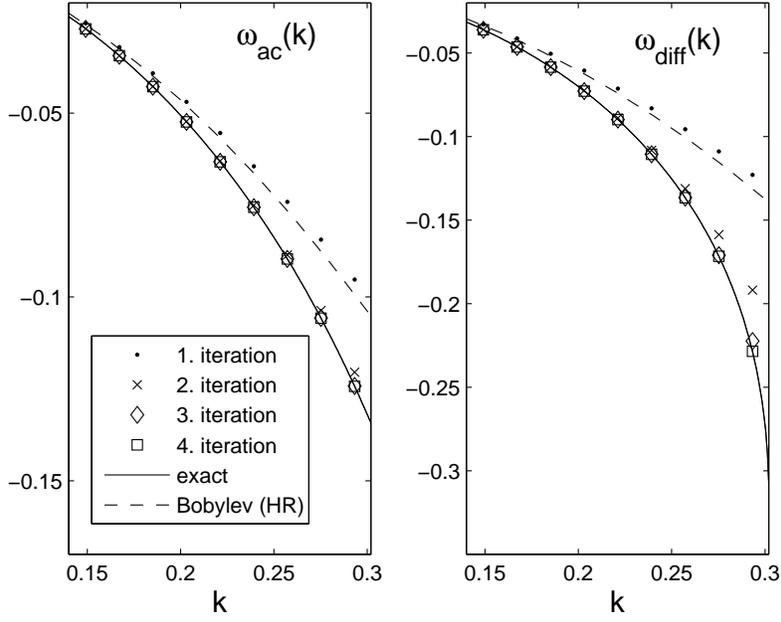}
\caption{Dispersion relations $\omega(k)$ for acoustic and
diffusive modes obtained via Newton iteration. In the plots, shown
is also the approximation obtained through Bobylev's hyperbolic
regularization (HR) \cite{Bo2006}. Newton iterations fail for
$k\geq k_c=0.3023$.} \label{mkfignewton}
\end{figure}

\section{Bobylev's hyperbolic regularization} \label{app:boby}

This appendix reviews a recent approach by Bobylev \cite{Bo2006}
and establishes a connection to the second-order variant of our
approach. We use the original notation of Ref.\  \cite{Bo2006} to
facilitate comparisons.

As was demonstrated in Ref.\  \cite{Bo2006}, after truncating the CE
expansion at the Burnett level, the (linearized) equation of
hydrodynamics takes the general form:
$\partial_{t}x+i(B_{0}+\varepsilon^2B_{1})x+\varepsilon
Ax+O(\varepsilon^3)=0$, where $x$ is the vector of hydrodynamics
variables $[\rho,u,T]$ and the operators $B_{0}$, $A$ and $B_{1}$
refer, respectively to Euler, Navier-Stokes and Burnett level of
approximation. $B=B_{0}+\varepsilon^2 B_{1}$ is a real non-symmetric
operator for $\varepsilon>0$. When applied to the Grad's system
(\ref{1b}), these findings are a special cases of equation
(\ref{short}) with (\ref{mats}) upon identifying $B_n=(-1)^n
\IM{n}$, $A=A_0$ and $A_n = (-1)^n \RE{n}$. The loss of symmetry of
the operator $B$, was identified as the reason of the instability
occurring in the Burnett equations. In order to cure this loss of
symmetry, HR introduces a symmetric a real valued operator $R$ and
defines a change of variables such that $z=x+\varepsilon^{2}Rx$ --
or in our notation above, ${\bf T}=({\bf 1}+\varepsilon^{2}R)$.
Hence, the resulting equation of hydrodynamics attains the form:
$z_{t}+i[B_{0}+\varepsilon^2(B_{1}+RB_{0}-B_{0}R)]z+\varepsilon
Az+O(\varepsilon^3) = 0$, more generally ${\bf z} = {\bf T}{\bf
M}{\bf T} ^{-1}{\bf z}$.
The suggested regularization consists in writing $\bf T^{-1}$ as a
polynomial (Taylor) expansion in powers of $\varepsilon$ and in
truncating it, as for $\bf T$, at second order.

HR, therefore, provides a regularization which is exact up to the
order $\varepsilon^2$. The operator $R$ has to be chosen in such a
way that $\widetilde{B}_{1}=B_{1}+[R,B_{0}]$ is real and symmetric,
where $[R,B_{0}]\equiv RB_{0}-B_{0}R$. It is instructive to consider
the HR as applied to the example of the 1D13M (\ref{1b}) which has
originally been written, in matrix notation, as:
\begin{eqnarray}
\partial_{ t} \left(\begin{array}{c}
                             \rho_{k} \\
                             u_{k}\\
                             T_{k}
                           \end{array}\right)  &=& \left\{- i\left[\left(\begin{array}{ccc}
                        0 & k & 0 \\
                        k & 0 & k \\
                        0 & \frac{2}{3}k & 0 \\
                      \end{array}
                    \right)+
                    \varepsilon^2\left(
                                            \begin{array}{ccc}
                                              0 & 0 & 0 \\
                                              \frac{4}{3}k^3 & 0 & -\frac{2}{3}k^3 \\
                                              0 & \frac{7}{6}k^3 & 0 \\
                                            \end{array}
                                          \right)\right] \right.  \nonumber \\
                                      && \left.    -\varepsilon\left(
                                                                  \begin{array}{ccc}
                                                                    0 & 0 & 0 \\
                                                                    0 & \frac{4}{3}k^2 & 0 \\
                                                                    0 & 0 & \frac{5}{2}k^2 \\
                                                                  \end{array}
                                                                \right)\right\}\left(\begin{array}{c}
                             \rho_{k} \\
                             u_{k}\\
                             T_{k}
                           \end{array}\right)+O(\varepsilon^{3}),
\label{toolong}
\end{eqnarray}
Expression (\ref{toolong}) offers those first terms of
(\ref{short}) with (\ref{mats}) for which the coefficients
($a_0=-4/3$, .. , $z_0=-7/4$) are analytically known, cf.
Tab.~\ref{tab1} for all values. Equation (\ref{toolong}) can hence
equivalently be formulated as ${\bf M} = \varepsilon \RE{0} - i (
\IM{0}-\varepsilon^2\IM{1} ) + O(\varepsilon^3)$. To apply the
regularization procedure to the system (\ref{1b}), one needs to
make matrix $B_{0}$ symmetric (it corresponds to restoring the
hyperbolicity of Euler equations, through a transformation ${\bf
T_{\alpha}}$). Then, introducing a real-valued, symmetric
(diagonal) matrix $R$ with diagonal elements $a(k)$, $b(k)$, and
$c(k)$ (which corresponds choosing a diagonal ${\bf T}$), and
imposing the symmetry of the resulting operator
$\widetilde{B}_{1}$ (more generally, of ${\bf T}\IM{n}{\bf
T}^{-1}$), the coefficients are interrelated as follows
\cite{Bo2006}
\begin{eqnarray}
  a(k) & = & b(k)+\frac{2}{3}k^2, \;\;\;
  c(k)  =
  b(k) - \frac{29}{24}k^2.
\end{eqnarray}
Notice, the transformation $ R_{0}R$ is a special case of
Eq.~(\ref{main}). The resulting operator $\widetilde{B}_{1}$ is
given by:
\begin{eqnarray}
\widetilde{B}_{1}&=&B_{1}+[R,B_{0}]=
B_{1}+b(k)[I,B_{0}]+[m_{ij},B_{0}] \nonumber \\
&=&B_{1}+[m_{ij},B_{0}],
\end{eqnarray}
and therefore unique (independent of $b(k)$). Hence, the
hydrodynamic equations resulting from HR as applied to 1D13M
attain the form:
\begin{eqnarray}
\partial_{ t} \left(\begin{array}{c}
                             \rho_{k} \\
                             u_{k}\\
                             T_{k}
                           \end{array}\right) = -\left(
                          \begin{array}{ccc}
                            0 & ik(1+\frac{2}{3}k^2\varepsilon^2) & 0 \\
                            ik(1+\frac{2}{3}k^2\varepsilon^2) & \frac{4}{3} k^2 \varepsilon &
                            \sqrt{\frac{2}{3}}ik(1+ \frac{13}{24} k^2\varepsilon^2 ) \\
                            0 & \sqrt{\frac{2}{3}}ik(1+ \frac{13}{24} k^2\varepsilon^2) &
                            \frac{5}{2}k^2 \varepsilon \\
                          \end{array}
                        \right)\left(\begin{array}{c}
                             \rho_{k} \\
                             u_{k}\\
                             T_{k}
                           \end{array}\right)+O(\varepsilon^3).
                           \label{eq14}
\end{eqnarray}

Since Eq.~(\ref{eq14}) is a special case of the more general
Eq.~(\ref{oursuper}), the connection to Bobylev's work has been explicitly established.

\end{appendix}

\end{document}